\documentclass[letter,dvipsnames,svgnames,x11names]{aa} 
\usepackage{graphicx}
\usepackage{soul}
\usepackage{txfonts}
\usepackage{natbib}
\usepackage[colorlinks=true, linktocpage, linkcolor={blue}, citecolor={blue}, urlcolor={blue}, backref=page]{hyperref}

\usepackage{siunitx}
\usepackage{bm}

\makeatletter
\newcommand{\subalign}[1]{%
  \vcenter{%
    \Let@ \restore@math@cr \default@tag
    \baselineskip\fontdimen10 \scriptfont\tw@
    \advance\baselineskip\fontdimen12 \scriptfont\tw@
    \lineskip\thr@@\fontdimen8 \scriptfont\thr@@
    \lineskiplimit\lineskip
    \ialign{\hfil$\m@th\scriptstyle##$&$\m@th\scriptstyle{}##$\hfil\crcr
      #1\crcr
    }%
  }%
}
\makeatother

\newcommand{\sigT}{\mbox{$\sigma_{\mbox{\tiny T}}$}}

\usepackage{color}

\begin{document}

   \title{GBT/MUSTANG-2 9\arcsec\ resolution imaging of the SZ effect in MS0735.6+7421}
   \subtitle{Confirmation of the SZ cavities through direct imaging}
   \titlerunning{MUSTANG-2 imaging of MS0735.6+7421}


\author{John Orlowski-Scherer
          \inst{1}
          \and
          Saianeesh K. Haridas\inst{1}
          \and
          Luca Di Mascolo\inst{2,3,4}
          \and
          Karen Perez Sarmiento\inst{1}
          \and
          Charles E. Romero\inst{5}
          \and
          Simon Dicker\inst{1}
          \and
          Tony Mroczkowski\inst{6}
          \and
          Tanay Bhandarkar\inst{1}
          \and
          Eugene Churazov\inst{7,8}
          \and
          Tracy E.\ Clarke\inst{9}
          \and
          Mark Devlin\inst{1}
          \and
          Massimo Gaspari\inst{10}
          \and
          Ian Lowe\inst{11}
          \and
          Brian Mason\inst{12}
          \and
          Craig L. Sarazin\inst{13}
          \and
          Jonathon Sievers\inst{14}
          \and
          Rashid Sunyaev\inst{7,8}
          }

\institute{Department of Physics and Astronomy, University of Pennsylvania, 209 South 33rd Street, Philadelphia, PA, 19104, USA\\\email{jorlo@sas.upenn.edu}
          \and
          Department of Physics, University of Trieste, via Tiepolo 11, 34131 Trieste, Italy
          \and
          INAF - Osservatorio Astronomico di Trieste, via Tiepolo 11, 34131 Trieste, Italy
          \and
          IFPU - Institute for Fundamental Physics of the Universe, Via Beirut 2, 34014 Trieste, Italy
          \and
          Center for Astrophysics | Harvard and Smithsonian, 60 Garden Street, Cambridge, MA 02143, USA
          \and
          European Southern Observatory (ESO), Karl-Schwarzschild-Strasse 2, D-85741 Garching, Germany
          \and
          Max Planck Institute for Astrophysics, Karl-Schwarzschild-Strasse 1, D-85741 Garching, Germany
          \and
          Space Research Institute (IKI), Profsoyuznaya 84/32, Moscow 117997, Russia
          \and
          Naval Research Laboratory, Code 7213, 4555 Overlook Ave SW, Washington, DC 20375 USA
          \and
          Department of Astrophysical Sciences, Princeton University, Princeton, NJ 08544, USA
          \and
          Department of Astronomy, University of Arizona, Tucson, AZ, 85721, USA
          \and
          NRAO, 520 Edgemont Rd, Charlottesville, VA, 22903, USA
          \and
          Department of Astronomy, University of Virginia, 530 McCormick Road, Charlottesville, VA, 22904-4325, USA
          \and
          Department of Physics, McGill University, 3600 University Street Montreal, QC, H3A 2T8, Canada
          }

   \date{Received July 20, 2022 accepted XX YY, 2022}

 
  \abstract
   {
   Mechanical feedback from active galactic nuclei is thought to be the dominant feedback mechanism quenching cooling flows and star formation in galaxy cluster cores. It, in particular, manifests itself by creating cavities in the X-ray emitting gas, which are observed in many clusters. However, the nature of the pressure supporting these cavities is not known. }
   {Using the MUSTANG-2 instrument on the Green Bank Telescope (GBT), we aimed to measure thermal Sunyaev-Zeldovich (SZ) effect signals associated with the X-ray cavities in MS0735.6+7421, a moderate-mass cluster that hosts one of the most energetic active galactic nucleus outbursts known.  We used these measurements to infer the level of nonthermal sources of pressure that support the cavities, such as magnetic fields and turbulence, as well as relativistic and cosmic ray components.
   }
   {We used the preconditioned gradient descent method to fit a model for the cluster, cavities, and central point source directly to the time-ordered data of the MUSTANG-2 signal. We used this model to probe the thermodynamic state of the cavities. }
   {We show that the SZ signal associated with the cavities is suppressed compared to the expectations for a thermal plasma with temperatures of a few tens of keV. The smallest value of the suppression factor, $f$, that is consistent with the data is $\sim$0.4, lower than what has been inferred in earlier work. Larger values of $f$ are possible once the contribution of the cocoon shock surrounding the cavities is taken into account.}
   {We conclude that in the ``thermal'' scenario, when half of the pressure support comes from electrons with a Maxwellian velocity distribution, the temperature of these electrons must be greater than $\sim100$\,keV at $2.5\sigma$ confidence.  Alternatively,
   electrons with nonthermal momentum distribution could contribute to the pressure, although existing data do not distinguish between these two scenarios.  The baseline model with cavities located in the sky plane yields a best-fitting value of the thermal SZ signal suppression inside the cavities of $f\sim0.5$, which, at face value,
   implies a mix of thermal and nonthermal pressure support. Larger values of $f$ (up to 1, i.e., no thermal SZ signal
   from the cavities) are still possible when allowing for variations in the line-of-sight geometry.}


   \maketitle
%


\section{Introduction}\label{sec:intro}
The majority of baryons in galaxy clusters reside in the diffuse intracluster medium (ICM). As baryons fall into clusters they are heated by shocks and compression, while simultaneously radiating away energy in the form of X-ray radiation \citep{Fabian1994}. Thus, in the absence of other processes, the cluster cores will radiate away all their heat in a short period of time. However, while X-ray observations of clusters reveal emission from the ICM, there is a notable deficit of soft X-rays, corresponding to temperatures of $\lesssim 1$\,keV \citep{Peterson2006}, as compared to predictions. 
One potential solution to this problem is that some process is injecting energy into the ICM and reheating it. Numerous mechanisms may provide this heating, but feedback by active galactic nuclei is believed to play the primary role \citep{2000A&A...356..788C,McNamara2007, McNamara2012,Gaspari2020, Hlavacek2022}. 

Jets are the main drivers of ICM reheating, although the exact mechanism is not clear yet. It is known that the jets, as traced by their synchrotron emission, often terminate in radio lobes that are coincident with depressions (cavities) in the X-ray emission \citep{Fabian2012}. The standard interpretation is that the jets form plasma bubbles in the ICM, a view that is supported by high-resolution hydrodynamical simulations \citep{Sternberg2009,Gaspari2011,Ehlert2019}. The nature of pressure support for these cavities (or radio bubbles) is poorly understood, and hence the means by which the energy from the jets couples to the ICM is also poorly understood. Broadly, the support mechanisms can be broken down into two categories: thermal and nonthermal. In the thermal support case, under the assumption of thermal equilibrium between the bubbles and the surrounding ICM, since the electron number density in the bubbles must be low (as evidenced by their low X-ray emission), the gas must be very hot to sustain the bubbles via pressure support. In the nonthermal case, the pressure might be due to a combination of relativistic protons, electrons, and magnetic fields.  

Observations of the thermal Sunyaev-Zeldovich (SZ) effect (\citealt{Sunyaev1970,Sunyaev1972}) provide a powerful complement to X-ray observations. 
Since the thermal SZ effect is sensitive to the line-of-sight integrated electron pressure of the ICM, it can distinguish the classical thermal pressure scenario from nonthermal pressure and relativistic electron populations \citep{Colafrancesco2003, Colafrancesco2005, Pfrommer2005, Mroczkowski2019, Yang2019}. Magnetic fields and ions do not contribute to the thermal SZ effect, while relativistic effects suppress the thermal SZ decrement. As a result, the thermal SZ signal from bubbles supported by nonthermal pressure will be suppressed. Conversely, thermally supported bubbles are relatively unsuppressed in the thermal SZ support scenario, 
unless the supporting particles are extremely hot (see Sect.~\ref{sec:sup_factor}).

MS 0735.6+7421 (hereafter MS0735) is a galaxy cluster at a redshift of $z=0.216$. It is notable for hosting two of the largest known X-ray cavities, nearly $200$\,kpc across, sourced by one of the strongest known radio outbursts in the Universe \citep{McNamara2005}. The immense size of these cavities allowed \citet{McNamara2005} to place strong constraints on the mechanical energy needed to create them, and hence the mechanical strength of the central radio source. This in turn firmly established the plausibility of radio outbursts as a mechanism for quenching cooling flows over long timescales. 

Recently, \citet[][hereafter A19]{Abdulla2019} put constraints on the pressure support of the bubbles in MS0735 using the Combined Array for Research in Millimeter-wave Astronomy (CARMA).  
As shown in, for example, \citet{McNamara2005}, \citet{Vantyghem2014}, \citet{Biava2021}, and \citet{begin2022}, the X-ray cavities in MS0735 correspond to radio-bright bubbles, indicating the presence of relativistic plasma. 
A19 found nearly complete suppression of the SZ signature of the bubbles at the $30$\,GHz observation frequency of the CARMA observations reported in \citet{Abdulla2019}; this implies nonthermal pressure support, or else thermal support by electrons with temperatures of at least $kT_{e} \gtrsim 150$\,keV. In this work, we build upon the work by A19 by observing MS0735 with the MUSTANG-2 instrument on the $100$\,m Green Bank Telescope (GBT). Observing at $90$\,GHz, MUSTANG-2 has comparable resolution ($\sim 9\arcsec$) to CARMA but higher sensitivity. 
   
The paper is structured as follows. An overview of the data is provided in Sect.~\ref{sec:data}. In Sect.~\ref{sec:analysis} we discuss the pressure profile fits across the cavities. We offer an interpretation in Sect.~\ref{sec:results}, and in Sect.~\ref{sec:discussion} we provide conclusions.
To facilitate comparison with A19, we adopt a $\Lambda$ cold dark matter concordance cosmology with $\Omega_\Lambda=0.7$ and $H_0=70~\rm km s^{-1} Mpc^{-1}$ throughout the work. At the redshift of the cluster ($z=0.216$), the 9\arcsec\ beam corresponds to 32~kpc.

\section{Data}\label{sec:data}


MUSTANG-2 is a $90$\,GHz bolometer camera on the GBT with $\sim 9\arcsec$ resolution and a $4.2\arcmin$ instantaneous field of view \citep{Dicker2014,Stanchfield2018}. 
The combination of resolution and field of view makes it well matched to MS0735, where the bubbles are $\sim 1\arcmin$ in diameter, as compared to the 9\arcsec\ beam, and the cluster profile has a characteristic radius of $\sim2\arcmin$ \citep[][hereafter V14]{Vantyghem2014}. 

Observations are saved as time-ordered data (TODs). 
To calibrate and preprocess the TODs, we used the MUSTANG-2 Interactive Development Language (IDL) 
pipeline MIDAS (\citealt{Romero2020}). The raw TODs are read and interpolated onto common timestamps and then calibrated. The calibration is obtained from regular  observations (every 20 minutes) of strong point sources interspersed with observations of the cluster. MIDAS then flags data from bad detectors as well as spikes due to, for example, glitches and cosmic rays.

\begin{figure*}
    \includegraphics[clip,trim=0.0cm 0.0cm 0.0cm 0.0cm,height=7.3cm]{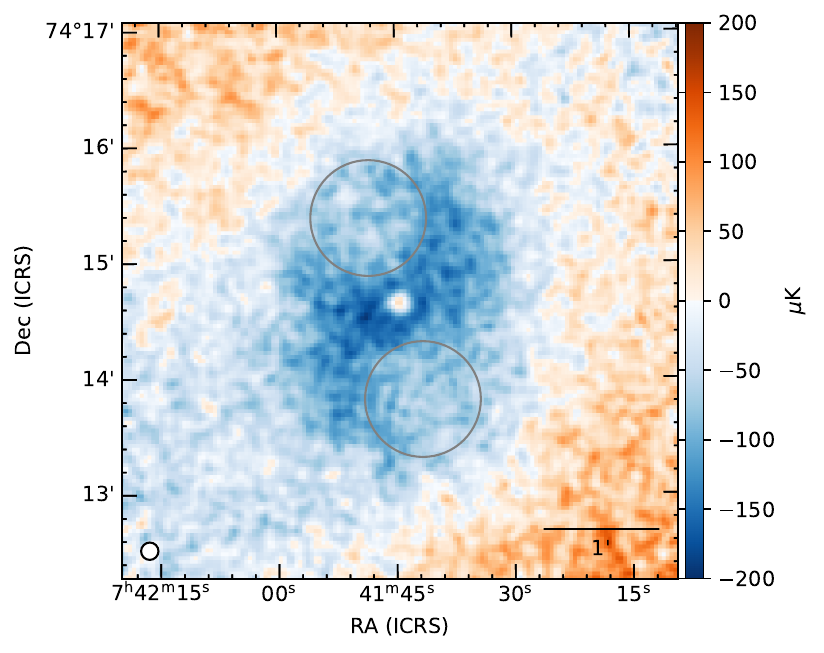}    
    \includegraphics[clip,trim=0.0cm 0.0cm 0.0cm 0.0cm,height=7.3cm]{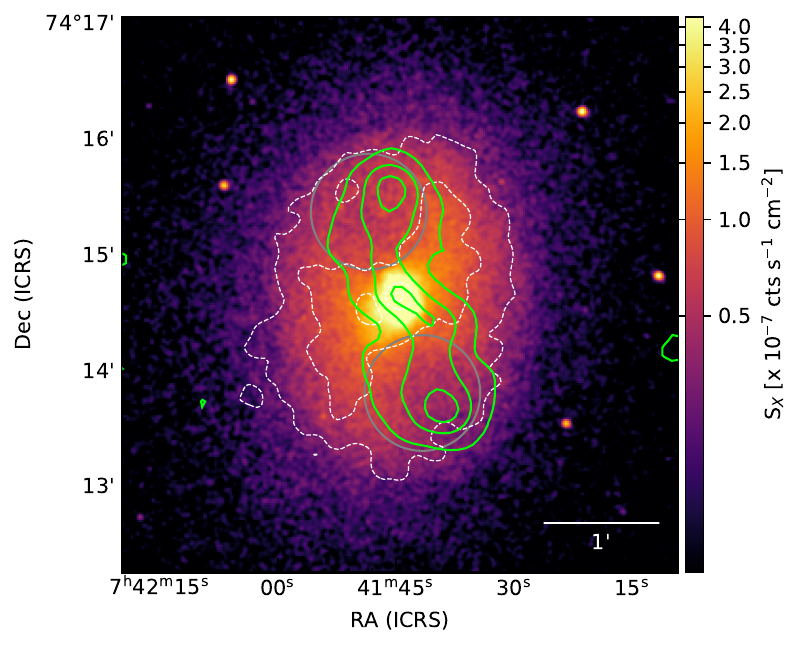}
    \caption{MS0735 in the SZ and X-ray.
    {\bf Left:} MUSTANG-2 image of MS0735.6+7421 in units of $\mu$K$_\textsc{cmb}$. The image is smoothed by a 1 pixel Gaussian (6\arcsec), yielding an effective resolution of 9.5\arcsec. 
    {\bf Right:} Exposure-corrected broadband (0.5-7.0~keV) {\it Chandra} X-ray image of MS0735.6+7421 using the same $\sim$500~ksec of data reported in V14.  The image is binned $2\times2$ pixels from the native resolution of 0.496\arcsec and smoothed by a 0.984\arcsec Gaussian.  Dashed white contours correspond to the signal to noise of the decrement seen in the MUSTANG-2 image (left), smoothed by 2 pixels, and are at a signal-to-noise ratio $\text{S/N}=-(1,2,3)$. The green contours correspond to $338$\,MHz continuum radio emission as traced by VLITE (resolution 21.5\arcsec $\times$ 16.2\arcsec) and trace the jets. At $z=0.216$, the 1\arcmin\ scale bar corresponds to 212~kpc.
    }
    \label{fig:sz+xray}
\end{figure*}

MUSTANG-2 spent 14 hours ($\sim 50.4$~ksec) observing MS0735 for projects AGBT21A\_123 and AGBT19A\_092. The resulting image is shown in the left panel of Fig. \ref{fig:sz+xray}.

\section{Analysis}
\label{sec:analysis}

To fit the data, we constructed a model comprising both the bulk pressure distribution and the bubble regions, as well as the central compact (point) source. We then fit this model directly to the TODs using the Preconditioned conjugate Gradient Descent (PGD) method as implemented in the Minkasi map-making code\footnote{\url{https://github.com/sievers/minkasi}}. The PGD method iteratively minimizes an objective function, in our case the likelihood of a model, by computing the gradient of that function and then ``stepping'' in the opposite direction of the gradient. The PGD method is computationally very fast and allows complicated models such as ours to be fit in a reasonable amount of time. In general, we followed A19, both in our choice of model and in the specific model parameters, where applicable. The model parameters themselves are generally derived from fitting to X-ray data as described in V14 (right panel of Fig.~\ref{fig:sz+xray}).  

\subsection{ICM model}
\label{sec:icm}

The distortion to the cosmic microwave background (CMB) intensity due to the thermal SZ effect is given in terms of the reduced frequency $x \equiv \frac{h\nu}{k_\textsc{b} T_\textsc{cmb}}$ \citep{Sunyaev1970} by\begin{align}
\label{eq:tsz_delta_I}
    \Delta I_{\nu} &\simeq I_0 \, y \, \frac{x^4 e^x}{(e^x-1)^2}\left( x\frac{e^x +1 }{e^x-1}-4\right)\nonumber\\
    &\equiv I_0 \, y \, g(x),
\end{align}
where
\begin{equation}
    I_0 = \frac{2(k_\textsc{b} T_{\textsc{cmb}})^3}{(hc)^2}
\end{equation}
and $g(x)$ encapsulates the spectral distortion of the thermal SZ effect, while $y = (\sigT/m_e c^2) \int P_e \, d\ell$ is called the Compton-$y$ parameter (see, e.g., \citealt{Carlstrom2002} and \citealt{Mroczkowski2019} for reviews).  Here, $P_e$ is the electron pressure, and $\ell$ is the path along the line of sight through the cluster. We note that at the MUSTANG-2 observing frequency, $90$\,GHz, the thermal SZ signal appears as a deficit in the CMB background.

Following A19, we modeled the global ICM pressure distribution in MS0735 as an elliptical double beta model, where a single beta model has the form\begin{equation}
    \label{eq:beta}
    P_{e} = P_{e,0} \left(1+\frac{x_1^2}{r_1^2}+\frac{x_2^2}{r_2^2}+\frac{x_3^2}{r_3^2}\right)^{\frac{-3\beta}{2}}
,\end{equation}
where $P_{e,0}$ is the pressure amplitude and $r_i$ are the core radii for each spatial direction. The profile is also rotated with respect to the standard right ascension (RA) and declination (Dec) coordinate axes. The predicted Compton-$y$ surface brightness is then given by Eq.~\ref{eq:tsz_delta_I}, 
where the integral is along the line-of-sight axis, $x_3$. 
The double beta model is then the sum of two beta models with different core radii, amplitudes, and betas. We assumed that the profiles have the same center RA and Dec, have the same ellipticity, and make the same angle with respect to the RA--Dec coordinate axis. As for the particular values of the model parameters, we used the X-ray-identified RA and Dec, the ellipticity and positional angle, and the X-ray profile exponents $\beta_1$ and $\beta_2$ from V14. Following A19, we derived $r_1$ and $r_2$ by requiring that their geometric mean be equal to the corresponding core radii from V14 and that their ratio equal the projected axis ratio from the same. The line of sight radius $r_3$ is not directly constrained by the available data.
As such, we considered two scenarios, one where $r_3 =r_1$ and another where $r_3=r_2$, which form an exploratory range for $r_3$. The amplitudes of the two beta models are free parameters. The chosen beta model values are summarized in Table~\ref{tab:iso_pars}.


\begin{table*}[]
    \centering    
    \caption{Summary of the ICM model parameters and their sources. The RA and Dec are the same for the two beta profiles. The superscript 1 denotes the outer beta profile, and the superscript 2 denotes the inner. Note that we set $r_3^1$ and $r_3^2$ to either the respective major or minor axis simultaneously. In other words, we did not set, e.g., $r_3^1 = r_2^1, r_3^2 = r_1^2$. $\theta$ is measured counterclockwise from the RA axis.}
    \begin{tabular}{cccc}
    \hline\hline\noalign{\smallskip}
        Parameter  & Source & Description  & Value  \\
        \hline
        RA & V14 & RA of MS0735& $07^{\text{h}} 41^{\text{m}} 44.5^{\text{s}}$ \\
        Dec & V14 & Dec of MS0735& $+74^{\circ} 14\arcmin 38.7\arcsec$  \\
        $r_1^1$ & V14 & {\footnotesize Semimajor axis of outer profile} & 341\,kpc \\
        $r_2^1$ & V14 & Semiminor axis of outer profile & 249\,kpc\\
        $r_3^1$ & A19 & Line-of-sight axis of outer profile & 249 or 341\,kpc \\
        $\beta_1$ & V14 & Slope of outer profile & 0.98 \\
        $A_1$ & N/A & Amplitude of outer profile & Free Parameter\\
        $r_1^2$ & V14 & Semimajor axis of inner profile & 167\,kpc \\
        $r_2^2$ & V14 & Semiminor axis of inner profile & 122\,kpc\\
        $r_3^2$ & A19 & Line-of-sight axis of inner profile & 122 or 167\,kpc \\
        $\beta_2$ & V14 & Slope of inner profile &  0.98 \\
        $A_2$ & N/A & Amplitude of inner profile & Free Parameter\\
        $\theta$ & V14 & Angle of MS0735 & $97^{\circ}$\\
        \hline
    \end{tabular}

    \label{tab:iso_pars}
\end{table*}


\subsection{Compact source}
A radio-bright compact source lies at the center of MS0735, corresponding to the active galactic nucleus itself. At the 9\arcsec\ resolution of MUSTANG-2, the source is unresolved, and we modeled it as a point source. We first fit the point source, treating the position (RA and Dec), half width, and amplitude as free parameters. Then we performed the full joint fit of the ICM profile, bubbles, and point source, fixing the RA and Dec and half width of the point source to the values found previously, but keeping the amplitude as a free parameter. The final, fixed values for the RA and Dec were RA = $07^{\text{h}} 41^{\text{m}} 44.6^{\text{s}}$ and Dec = $+74^{\circ} 14\arcmin 39.3\arcsec$, and the half width was $2\arcsec7$.







\subsection{Radio lobe emission}
Contamination of the SZ signal by radio emission is a distinct possibility, especially as cavities are frequently coincident with the radio lobes. The radio emission associated with M0735 is shown in the right panel of Fig.~\ref{fig:sz+xray}, where we overlay 338 MHz radio contours from the VLA Low-band Ionosphere and Transient Experiment \citep[VLITE;][]{Clarke2016,Polisensky2016} on the X-ray data. The jets are seen to terminate at the location of the cavities. To assess the potential risk of contamination, we followed A19 in estimating the flux using a power-law spectrum, $S\propto \nu ^{-\alpha}$. The VLA-measured flux densities were $720$\,mJy at $327$\,MHz and $11.7$\,mJy at $1.4$\,GHz within the lobes \citep{Birzan2008}; fitting these two points to a power law yields a spectral index of $\sim 2.8$ and an estimated flux at $90$\,GHz of $0.08\,\mu$Jy, far below our noise level ($\sim 10\mu$\,K). From earlier observations, \citet{Cohen2005} found lobe emission that was about twice as  high but a similarly steep spectral index that also made a negligible contribution. Consequently, we did not include radio lobe emission in our model. 

\subsection{Shock}
MS0735 is known to have an elliptical shock front. We modeled the cocoon shock as an enhancement of the pressure within the shock, including within the bubbles, by a uniform amount. This amount is parameterized by the Mach number, $\mathcal{M}$, which was a fit parameter. We took the shock geometry from V14. To confirm the SZ detection of the shock, we performed fits both with and without the shock enhancement.

\subsection{The bubbles}
\label{sec:sup_factor}
We treated the bubbles by taking the geometry from the X-ray data. We assumed that the SZ signal within the bubbles is uniformly suppressed by some factor $f$; that is, if the Compton-$y$ signal for the double beta profile is given by $h(x,y,z)$, then inside the bubble it is given by $(1-f)h(x,y,z)$. From the X-ray data, we approximated the bubbles as spherical in shape with radius $100$\,kpc $\simeq 30\arcsec$ (V14). When calculating the model, we multiplied the signal for all points within the bubbles by a suppression factor $0\leq f \leq 1$, which is a free parameter of the model and allowed us to differentiate between the bubbles. We note that the positions of the bubbles along the line of sight are unknown; we assumed them to be in the plane of the sky (see Sect.~\ref{sec:discussion} for a discussion of the effect of moving the cavities out of the plane of the sky). The bubble parameters are summarized in Table~\ref{tab:bubble_pars}.


\begin{table}[b!]
    \centering    
    \caption{Fixed coordinates and radii for the two bubbles. The values were taken from V14.}
     
    \begin{tabular}{cccc}
    \hline\hline\noalign{\smallskip}
        Bubble  & $\Delta$RA   & $\Delta$Dec & radius   \\
        
                      &  $\arcsec$    & $\arcsec$    &    $\arcsec$ \\   
        \hline
        Northeast & -15 & 43 &30\\
        Southwest & 21 & -51 &30\\\hline

    \end{tabular}

    \label{tab:bubble_pars}
\end{table}

\begin{figure}[h]
    \centering
    \includegraphics[width=\columnwidth]{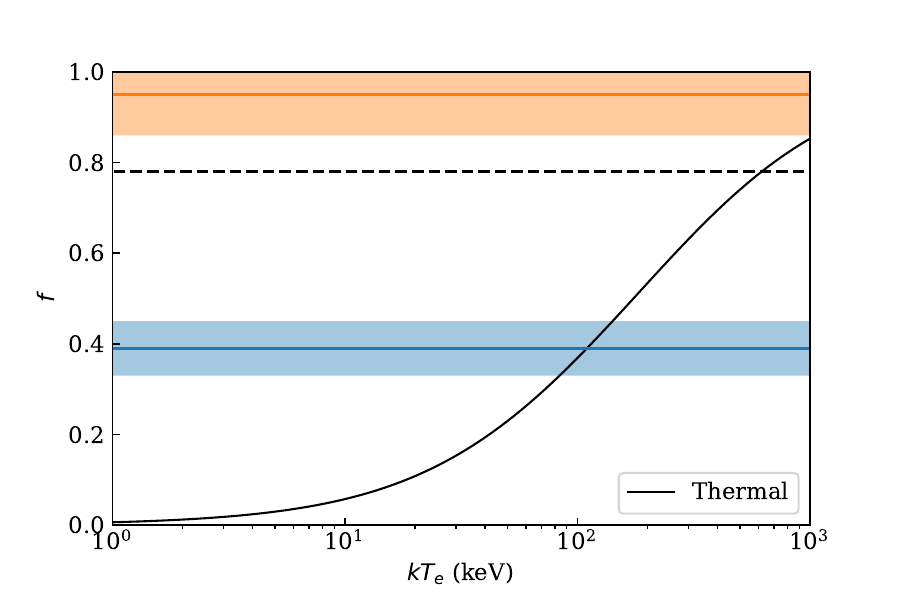}
\caption{Plot of the suppression factor $f$ vs. kT for the thermal support case. $f =1$ means complete suppression, i.e., no SZ signal from the bubbles, while $f=0$ means no suppression, i.e., the signal within the bubble is identical to the global ICM signal. The blue band shows the best fit $f$ with $1\sigma$ uncertainties for the lowest $f$ case considered in Sect.~\ref{sec:results}, corresponding to thermal pressure support by electrons with temperatures of at least $110$\,keV. The dashed line shows the lowest value consistent with \cite{Abdulla2019} to $1\sigma$.}
\label{fig:sup_factor}
\end{figure}

In order to interpret these suppression factors, we had to convert them into electron temperatures. In both the thermal and nonthermal scenarios discussed in Sect.~\ref{sec:intro}, the electrons contribute to the SZ signal. We therefore wanted to calculate the expected SZ signal, $\Tilde{g}(T,x)$, in these two scenarios and compare them to the expected signal in the bubble regime if the ICM were unperturbed, $g(T,x)$, where both spectra are functions of the electron temperature, $T$, and the reduced frequency, $x\equiv \frac{h \nu}{k_\textsc{b} T_\textsc{cmb}}$. The suppression factor is then $f = 1 - \frac{\Tilde{g}}{g}$. 
The full derivation is given in Appendix~\ref{ap:f}. The results of this derivation, which is $f$ as a function of temperature in the thermal case and lower momentum cutoff in the nonthermal case, are shown in Fig.~\ref{fig:sup_factor}.


If the suppression factor is near $0$, then the signal from the bubbles is consistent with the global ICM profile, that is to say, the gas in the bubbles is in thermal pressure equilibrium.  When the suppression factor is near one, then there is no SZ signal from the bubble, which in turn implies either that the support is nonthermal or that the electrons in the bubbles are very hot ($\gtrsim 100$\,keV).


\begin{table*}[tbh!]
    \centering    
    \caption{Summary of the results of the various fitting routines we completed. ``TOD subtract'' indicates whether the estimated elevation synchronous signal was subtracted from the data or not (see Sect.~\ref{sec:results}). $\beta_1$ is the power law for the outer beta profile: if no uncertainty is given, then it was fixed in that model; if an uncertainty is given, then it was a free parameter. $\mathcal{M}$ is the Mach number; if it is 0, then the shock was not included in that fit. The column $r_3$ indicates whether the line-of-sight core radius was set to the semimajor ($r_1$) or semiminor ($r_2$) core radius. $f_{NE}$ and $f_{SW}$ are the suppression factors for the northeast and southwest bubbles, respectively. $\text{T}_{NE}$ and $\text{T}_{SW}$ are the implied temperatures in the bubbles assuming full pressure support; it is the temperature implied by $f$ as shown in Fig.~\ref{fig:sup_factor}. For each model, an F-test was performed between that model and the same model without cavities. The significance of this test is reported in the last column.}
    \begin{tabular}{ccccccccc}
    \hline\hline\noalign{\smallskip}
        TOD Subtract   & $\beta_1$ & $\mathcal{M}$ & $r_3$ & $f_{NE}$  & $f_{SW}$ & $\text{T}_{NE}$ (keV) & $\text{T}_{SW}$ (keV) & Significance \\
        \hline
        Yes & $1.29\pm 0.06$ &               0 & $r_1$ & $0.95\pm 0.09$ & $0.74\pm 0.10$ & $3100\substack{+3000\\-1000}$ & $500\substack{+900\\-100}$ & 8.31\\
        Yes & $1.32\pm 0.06$ &               0 & $r_2$ & $0.73\pm 0.07$ & $0.57\pm 0.08$ & $475\substack{+225\\-175}$    & $250\substack{+100\\-50}$   & 8.35\\
        No  & $0.97\pm 0.04$ &               0 & $r_1$ & $0.79\pm 0.08$ & $0.51\pm 0.08$ & $650\substack{+500\\-225}$   & $175\substack{+75\\-50}$   & 8.43\\
        No  & $1.10\pm 0.04$ &               0 & $r_2$ & $0.61\pm 0.06$ & $0.40\pm 0.06$ & $275\substack{+75\\-75}$    & $115\substack{+30\\-25}$   & 8.17\\
        Yes &           0.98 &               0 & $r_1$ & $0.81\pm 0.09$ & $0.60\pm 0.10$ & $750\substack{+800\\-300}$   & $400\substack{+250\\-150}$  & 7.12\\
        Yes &           0.98 &               0 & $r_2$ & $0.62\pm 0.07$ & $0.46\pm 0.08$ & $275\substack{+100\\-75} $    & $ 150\substack{+50\\-50}$  & 7.02\\
        No  &           0.98 &               0 & $r_1$ & $0.80\pm 0.07$ & $0.52\pm 0.08$ & $700\substack{+450\\-225}$   & $200\substack{+75\\-50}$   & 8.52\\
        No  &           0.98 &               0 & $r_2$ & $0.61\pm 0.06$ & $0.39\pm 0.06$ & $275\substack{+75\\-75}$     & $110\substack{+30\\-25}$    & 8.41\\
        Yes &           0.98 & $1.78  \pm 0.14$  & $r_1$ & $0.93\pm 0.09$ & $0.71\pm 0.10$ & $2250\substack{+3000\\-1325}$   & $425\substack{+300\\-150}$  & 8.49\\
        Yes &           0.98 & $1.82  \pm 0.14$  & $r_2$ & $0.77\pm 0.07$ & $0.61\pm 0.08$ & $575\substack{+325\\-175}$   & $275\substack{+75\\-75}$   & 8.61\\
        No  &           0.98 & $1.15 \pm 0.05$ & $r_1$ & $0.84\pm 0.07$ & $0.55\pm 0.08$ & $900\substack{+800\\-325}$   & $225\substack{+75\\-50}$   & 8.90\\
        No  &           0.98 & $1.20 \pm 0.05$ & $r_2$ & $0.67\pm 0.06$ & $0.45\pm 0.06$ & $350\substack{+150\\-100}$     & $140\substack{+40\\-30}$   & 8.99\\
        \hline

    \end{tabular}

    \label{tab:ms0735_res}
\end{table*}

\subsection{Bowling}
Residual elevation-dependent noise has been observed in some MUSTANG2 data, which we refer to as ``bowling.'' As an object moves throughout the course of an observation, this noise is essentially rotated about the center of the observation, converting the elevation-dependent noise into a radial gradient. This leads to large-scale features on the order of the size of the maps (>8'). This effect has been observed before in MUSTANG-2 data \citep{Dicker2020}. Similar to \cite{Dicker2020}, we offset some pointings from the cluster center. In the case of the AGBT19A\_092 observations, some pointings were offset to the south, while in AGBT21A\_123 we followed \cite{Dicker2020} in using a mix of central pointings as well as four pointings offset by $1.5\degr$. 
However, this was not able to completely remove the bowling. To further mitigate it, we then fit a second-order polynomial to the elevation versus the signal 
for each TOD and estimated the elevation synchronous signal. We then subtracted this polynomial from the data before estimating the noise. We fit our model to the data both when this TOD subtraction was performed and when it was not. This is similar to the method used in \cite{Dicker2020} to remove residual bowling, the only difference being that we first subtracted the common mode from the TODs before fitting the second-order polynomial. In general, the results from the subtracted and un-subtracted TODs are in agreement. The bowling should also be down-weighted as noise in the fitting procedure via the noise estimation routines in Minkasi. The bowl's characteristic scale is the map scale, $\sim 6\arcmin$, and so it should not influence the parameter estimation for features near the center of the map with relatively smaller angular scales, such as the bubbles. 

\section{Results}
\label{sec:results}

Due to our inability to constrain the line-of-sight geometry of MS0735, we had to consider a variety of scenarios for that geometry. As discussed in Sect.~\ref{sec:icm}, we set the line-of-sight core radii, $r_3$, to equal either the semimajor ($r_1$) or semiminor ($r_2$) core radii. This provides an exploratory range for the suppression factors, bracketing the most extreme cases; in other words, both inner and outer profiles have $r_3 = r_1$ or both have $r_3 = r_2$. Similarly, we fit models both with and without enhancement of the SZ signal from the cocoon shock. Finally, we also considered both models where the outer profile slope, $\beta_1$, was fixed to the value from V14 (0.98) and ones where it was a free parameter of the model. We considered every permutation of TOD subtraction and $r_3=r_1$ or $r_2$. We could not fit for both $\beta_1$ and $\mathcal{M}$ simultaneously as the two parameters are degenerate within the shock envelope, and the observations lack the signal to noise outside the envelope to break that degeneracy.

\begin{figure}[h]
    \centerline{
    \includegraphics[width=\columnwidth]{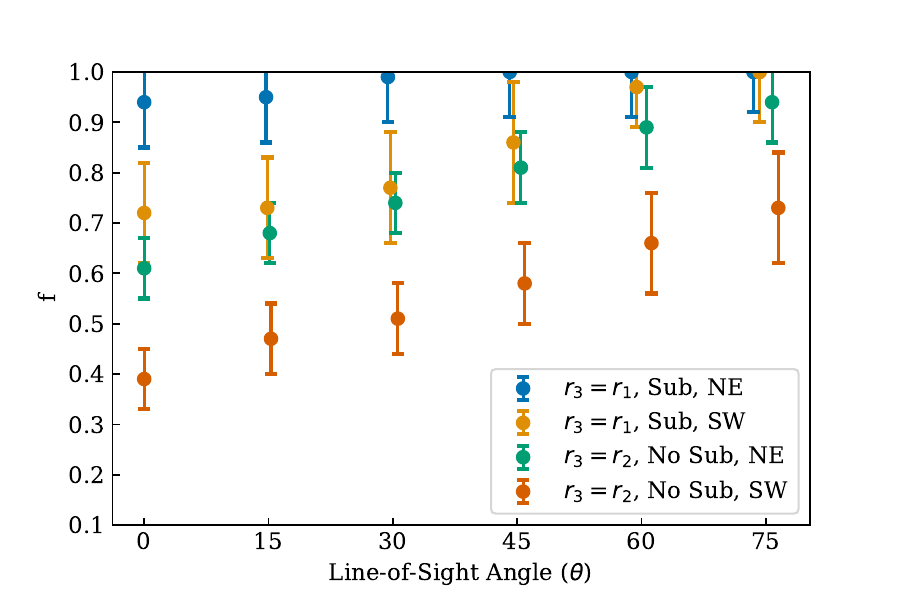}
}
\caption{Plot of the suppression factor, $f$, as a function of the line-of-sight angle, with $\theta = 0$ being in the plane of the sky and $\theta=90$ lying along the z-axis. Shown is the $f$ for both the northeast and southwest cavities for the scenarios in Table~\ref{tab:ms0735_res} that include shocks with the highest and lowest suppression factors. Explicitly, they are: with shock, $r_3 = r_1$, and with TOD subtraction; and with shock, $r_3 = r_2$, and without TOD subtraction.
In general, $f$ increases with increasing $\theta$, although we do not completely lose our ability to distinguish between pressure support scenarios, e.g., we can still rule out $f=1$ for the southwest cavity in the $r_3 = r_2$ without TOD subtraction.}
\label{fig:z_offset}
\end{figure}

We also investigated the effect of moving the bubbles along the line of sight. We took the permutations with shocks with the highest and lowest suppression factors (see Table~\ref{tab:ms0735_res}) and reran them with the bubbles offset at $15$, $30$, $45$, $60$, and $75\degr$ from the plane of the sky. The results are shown in Fig.~\ref{fig:z_offset}. In general, the suppression increases with increasing angle. This makes sense, as moving the bubble along the line of sight moves it into more tenuous areas of the ICM. As a result, the integrated pressure of the bubble is lower, and thus the suppression within the bubble must be higher to produce the same effect. The effect ranges up to a $60$\% increase in $f$ for the most extreme angles, although a $20$\% increase is typical for more moderate angles. While this does not completely degrade our ability to distinguish between support scenarios (our lowest suppression factor is still inconsistent with 1 at $\sim 4 \sigma$), it does reinforce the need for multiwavelength SZ observations to disentangle the effects of different pressure support scenarios from the effect of line-of-sight geometries. Of note, for this plot we enforced $f \leq 1$; we also ran fits without that enforcement. The suppression factor remained consistent with $f \leq 1$ within uncertainties. Had it not, it would have indicated that some of our geometrical assumptions, either about the bubbles or the ICM profile, were incorrect.

\begin{figure*}[bth!]
    \centerline{
    \includegraphics[clip,trim=0.0cm 0.0cm 0.0cm 0.0cm,height=0.27\textwidth]{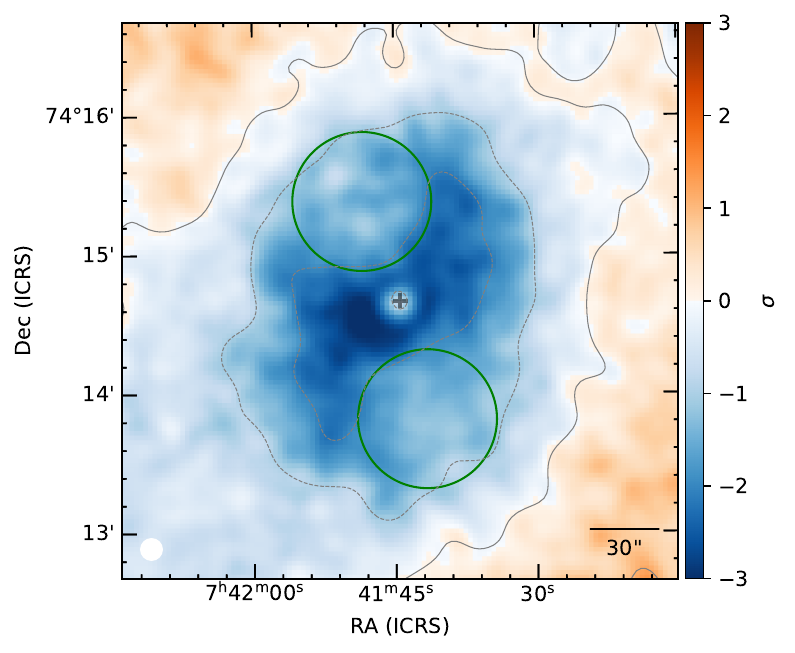}
    \includegraphics[clip,trim=0.0cm 0.0cm 0.0cm 0.0cm,height=0.27\textwidth]{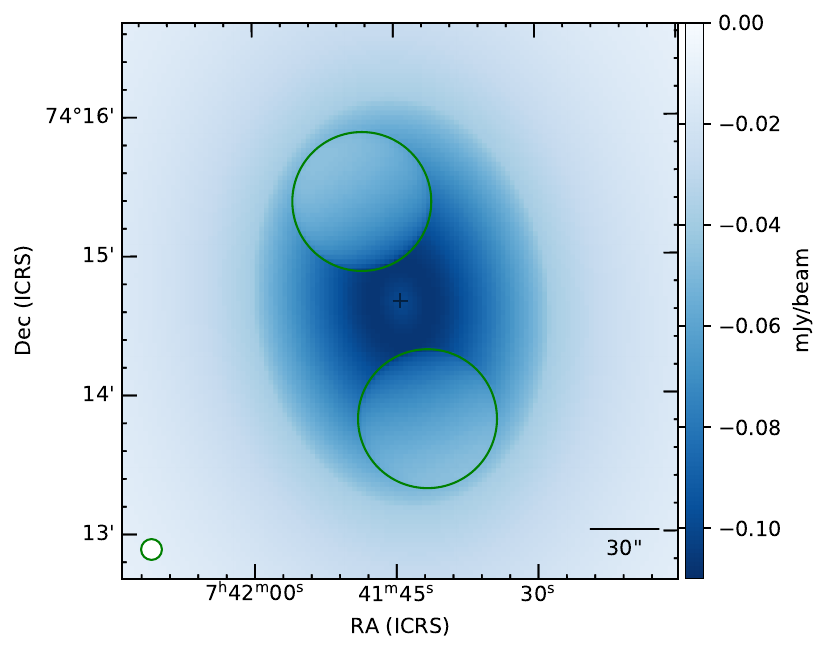}
    \includegraphics[clip,trim=0.0cm 0.0cm 0.0cm 0.0cm,height=0.27\textwidth]{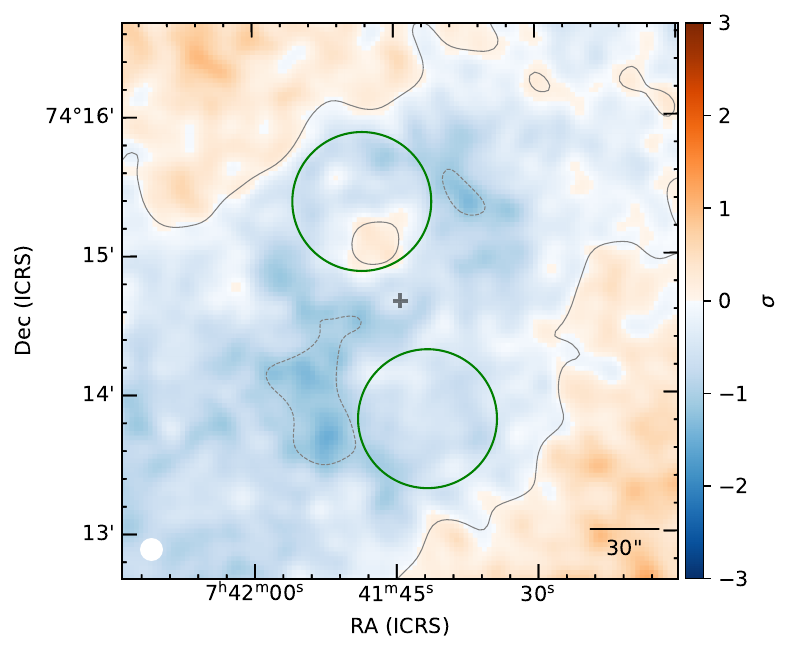}
    }
\caption{Signal-to-noise ratio (S/N) map for MS0735 for the data (left) and residual of the data minus the TOD with shock $r_3=r_1$ variation (right), which is the ninth variation listed in Table~\ref{tab:ms0735_res}. The contours are at steps of $\text{S/N} =1$. The cross indicates the cluster center, and the circles show the X-ray-identified bubble locations. The white circle in the bottom left shows the MUSTANG-2 beam. The noise at the center of the maps is $\sim 10\mu$K.
}
\label{fig:ms0735_resid}
\end{figure*}

The results of these fits are given in a systematic way in Table~\ref{tab:ms0735_res}, and a plot of the data and the residuals of the data for one of our models is shown in Fig.~\ref{fig:ms0735_resid}. The left plot shows the MUSTANG-2 observations of MS0735. 
The right plot shows the residuals of the data with the TOD subtracted, with shock, $r_3=r_1$ variation. Of all these permutations, the most directly comparable to A19 are those with $\beta_1 = 0.98$ and without TOD subtraction or the shock enhancement. These should be considered the baseline, while the permutations serve as a consistency check.

We consistently find higher suppression factors in the northeast bubble as compared to the southwest bubble. The $f_{NE}$ values range from $0.61 - 0.95$, while $f_{SW}$ ranges from $0.39 - 0.74$. For both the northeast and southwest bubbles, the resulting suppression factors indicate that if the pressure support in the bubbles is thermal, it must be coming from electrons with temperature $\gtrsim 100$\,keV. Including the shock raises the suppression factors by about $1\sigma$ with respect to models without the shock.







To quantify the success of our model, for each variation presented in Table~\ref{tab:ms0735_res} we fit a model in the same way with the same parameters, except with the bubble suppression fixed to $f=0$. 
For each variation, we then performed an F-test between the variation and its corresponding no-bubble model. We list the significances in Table~\ref{tab:ms0735_res}. They should not be used to select between the models; it is only to show that all variations significantly improve the fit as compared to the same model without bubbles.

\section{Discussion and conclusions}
\label{sec:discussion}
For all of the variations we considered, support by thermal electrons with temperatures $\lesssim 10$\,keV ($f\simeq 0.06$) is excluded by at least $4.5\sigma$. The lowest possible suppression factor at $2\sigma$ is $f= 0.39 - 2 \cdot 0.06 = 0.27$, roughly corresponding to thermal support from electrons at $\sim 60$\,keV; this limit is only for the southwest bubble. For the northeast bubble, pressure support by thermal electrons with a temperature of $100$\,keV ($f\simeq 0.37$) is excluded at $4\sigma$, while the lowest temperature not excluded at $2\sigma$ is $\sim 200$\,keV. 

While our best fit suppression factors were significantly lower than those found in A19, our findings still support their general conclusions that if the bubbles in MS0735 are supported by thermal pressure, the plasma in the northeast bubble must be at least\footnote{Note the conversion from suppression factor to temperature is highly nonlinear, so while, e.g., the $1\sigma$ constraint is 60\,keV, the $2\sigma$ constraint is not 120\,keV but 125\,keV.} 
$\geq 280\substack{+80\\-60}$\,keV. Alternatively, the cavities may be supported by particles with a nonthermal momentum distribution, or the support may be provided by magnetic fields \citep{Braithwaite2010} or by turbulence or dynamical pressure \citep[e.g.][]{Wittor2020}. A broad range of nonthermal pressure support mechanisms have the potential to suppress the thermal SZ signal, including up to complete suppression ($f=1$). Of course, the work here cannot rule out a combination of thermal pressure support with other sources of pressure support. 

In general, when we fixed our outer beta to the value reported in V14, we recovered lower suppression factors than when we made $\beta_1$ a fit parameter. When $\beta_1$ was a free parameter, we favored steeper values than V14, and, correspondingly, our fit amplitude was higher. 
This leads to the outer beta profile being $\sim 20\%$ larger in amplitude at the radius of the bubbles, and correspondingly requires a higher suppression factor to fit the data. 
Since the free $\beta$ model is a strict superset of the fixed $\beta$ model, we could again apply an F-test to determine whether the fit values of $\beta$ are preferred over the V14 at a statistically significant level. These results can be found in Table~\ref{tab:beta_sig}. In all cases, we find statistically significant support for the higher fit values of $\beta$. The significance is stronger when the TODs have been bowling-subtracted, which may indicate some degree of degeneracy between the outer profile $\beta$ and the bowling effect. 
As such, and to have a direct comparison to A19, we report fits with both a fixed and fit $\beta$.


\begin{table}[]
    \centering    
    \caption{Statistical significance of the improvement of fit as determined by an F-test for freeing the outer slope, $\beta_1$, for various combinations of TOD subtraction and $r_3$ values. In general, the fit is improved at a statistically significant level when performing TOD subtraction, but did not improve without it. This is indicative of degeneracy between the bowling of the maps and $\beta_1$. See the discussion in Sect.~\ref{sec:discussion}.}
    
    \begin{tabular}{cccccc}
    \hline\hline\noalign{\smallskip}
        TOD Subtract & $\beta_1$ & $r_3$ &  Significance \\
        \hline
        Yes & $1.29\pm 0.06$ & $r_1$ & 7.06\\
        Yes & $1.32\pm 0.06$ & $r_2$ & 8.00\\
        No & $0.97\pm 0.04$ & $r_1$ & 0 \\
        No & $1.10\pm 0.04$ & $r_2$ & 0\\
        \hline

    \end{tabular}

    \label{tab:beta_sig}
\end{table}


Similarly, we computed the significance of our shock detection via an F-test comparison with the corresponding no-shock model. The results are given in Table~\ref{tab:shock_sig}. With the TOD subtraction, the shock detection is significant at the $\sim 13\sigma$ level, while without the subtraction it is not significant. This may be because the bowling is of comparable scale ($\sim 3\arcmin$) to the shock front.


\begin{table}[]
    \centering    
    \caption{Statistical significance of the improvement of fit as determined by an F-test for adding the shock enhancement for various combinations of TOD subtraction and $r_3$ values. The inclusion is very statistically significant when TOD subtraction is performed, but marginal when it is not. This may be because the bowling is of comparable scale to the shock, and hence without TOD subtraction we have difficulty detecting the shock. }
    
    \begin{tabular}{cccccc}
    \hline\hline\noalign{\smallskip}
        TOD Subtract & $\mathcal{M}$ & $r_3$ &  Significance \\
        \hline
        Yes & $1.78\pm 0.14$ & $r_1$ & 9.91\\
        Yes & $1.82\pm 0.14$ & $r_2$ & 11.40\\
        No & $1.15\pm 0.05$ & $r_1$ & 3.60 \\
        No & $1.20\pm 0.05$ & $r_2$ & 5.32\\
        \hline

    \end{tabular}

    \label{tab:shock_sig}
\end{table}


While in general our models, due to the assumptions made in our analysis,  do not require complete suppression of the SZ signal within the bubbles, we cannot rule it out. For example, if the line-of-sight core radius were to be larger than the semimajor axis of either of the axes in the plane of the sky, 
then the actual suppression would be higher than the highest results presented here. Similarly, our assumption that the bubbles are in the plane of the sky places them at the maximum possible SZ signal given their angular location. If they are not in the plane of the sky, then the integrated SZ signal from the bubbles would be lower than assumed by our model, and hence our fit suppression factor will be biased low.


In this work we have demonstrated the capability of MUSTANG-2 to constrain the thermal content of cavities in the ICM of a cluster. While the typical cavity is significantly smaller than those in MS0735 ($\sim 2-10\arcsec$ in \citealt{Hlavacek2015}), other clusters with large cavities would likely also prove to be good candidates for MUSTANG-2 observations.  
Looking to the near future, the upcoming TolTEC 
experiment \citep{Wilson2020} will undertake observations of clusters with cavities with sufficient resolution ($\sim 5\arcsec$) to resolve many cavities. TolTEC will also provide a multi-chroic view of clusters, which may prove useful for distinguishing between support mechanisms \citep{Colafrancesco2003, Colafrancesco2005}. 
Farther south,  Atacama Large Millimeter submillimeter Array (ALMA) continues to provide the potential to observe cavities, with the caveat that they have (subarcminute) scales accessible after interferometric filtering by ALMA; in the longer-term, upcoming and proposed facilities such as Square Kilometre Array (SKA) and the Atacama Large Aperture Submillimeter Telescope \citep[AtLAST;][]{Klaassen2020} will provide a more complete view. Specifically, future high-resolution observations spanning $\nu \sim 30-500$~GHz will able to obviate the geometrical effects discussed above by directly probing the full SZ spectrum of the bubbles.

\begin{acknowledgements}
MUSTANG-2 is supported by the NSF award number 1615604 and by the Mt.\ Cuba Astronomical Foundation. 
This material is based upon work supported by the Green Bank Observatory. 
GBT data were acquired under the project IDs AGBT21A\_123 and AGBT19A\_092.
The Green Bank Observatory is a facility of the National Science Foundation operated under cooperative agreement by Associated Universities, Inc. The National Radio Astronomy Observatory is a facility of the National Science Foundation operated under cooperative agreement by Associated Universities, Inc. This research was enabled in part by support provided by SciNet (\url{https://www.scinethpc.ca/}) and Compute Canada (\url{https://www.computecanada.ca}). The scientific results reported in this article are based on observations made by the
Chandra X-ray Observatory. Basic research in Radio Astronomy at the Naval Research Laboratory is funded by 6.1 Base funding. Construction and installation of VLITE was supported by the NRL Sustainment Restoration and Maintenance fund. Massimo Gaspari acknowledges partial support by NASA Chandra GO9-20114X and HST GO-15890.020/023-A, and the {\it BlackHoleWeather} program.
We thank the referee for comments that helped improve the work presented.

\end{acknowledgements}

\bibliographystyle{aa}
\bibliography{ms0735}

\appendix

\section{The suppression factor}
\label{ap:f}
The derivation of the suppression factor in the thermal 
case follows closely from  A19, \citet{Colafrancesco2003}, and \citet{Ensslin2000}.
Given the number density of electrons, $n_e$, in the cavities, the optical depth of the cavities is
\begin{equation}
    \label{eq:tau_cav}
    \tau_{cav} = \sigma_T \int_{cav} n_e d\ell
,\end{equation}
where the subscript $cav$ indicates that the integration is performed over the cavity and $d\ell$ indicates that it is along the line of sight. We can then determine the change in SZ flux density by considering the difference $\delta i(x) = j(x)\tau_{cav} - i(x)\tau_{cav}$. Here, $i(x)$ is the Planck distribution, and its product with $\tau$ determines the scattering of photons from $x$ to other frequencies, while $j$ governs the scattering of photons to $x$ from other frequencies. From \citet{Ensslin2000}, $j$ is given by\begin{equation}
    j(x) = \int_0^{\inf} dt \int_0^{\inf} P(t;p) i(x/t) f_e(p) dp
,\end{equation}
where $P(t;p)$ is the photon redistribution function for a mono-energetic electron distribution and $f_e(p)$ is the electron momentum spectrum. We used the analytic form of $P(t;p)$ derived in the appendix of \citet{Ensslin2000}. We can now rewrite $\delta i(x)$ in terms of the the SZ spectral shape, $g(x)$, and the spectrum for the electron distribution within the cavity as 
\begin{equation}
    \delta i(x) = [j(x) -i(x)]\tau_{cav} = y_{cav} \Tilde{g}(x)
,\end{equation}
where $y_{cav}$ is the amplitude of the Compton-y in the cavity, which is given by \cite{Ensslin2000}:
\begin{equation}
    y_{cav} = \frac{\sigma}{m_e c^2}\int n_e k \Tilde{T}_e d\ell
\end{equation}
for $k\Tilde{T}_e = \frac{P_e}{n_e}$. We note that $P_e$ is the electron pressure, not the electron redistribution function. We can then further write an expression for $\Tilde{g}(x)$:
\begin{equation}
    \Tilde{g}(x) = [j(x) -i(x)]\frac{m_ec^2}{\langle k\Tilde{T}_e\rangle}
,\end{equation}
where $ k\Tilde{T}_e$ is the pseudo-temperature, which is equal to the temperature in the thermal case and is defined by $\langle k\Tilde{T}_e\rangle = \frac{\int  n_e  k\Tilde{T}_e d\ell}{\int n_e d\ell}$. 

Now we can write an equation for the suppression factor, $f$. Following the assumptions established above -- that the cavities are spherical and centered on the plane of the sky, and that the Compton-y profile in the bubbles is the same as outside modulo the suppression factor -- we have
\begin{equation}
    \delta i(x) = [y_{cl} - y_{cav}]g(x) + y_{cav}\Tilde{g}(x)
.\end{equation}
Factoring out $g(x)$ and defining $f\equiv 1 - \frac{\Tilde{g}(x)}{g(x)}$,
\begin{equation}
    \delta i(x) = (y_{cl} - fy_{cav})g(x)
.\end{equation}
We can now connect our observed suppression factor, $f$, to the underlying electron temperature via the modified Compton-y spectrum, $\Tilde{g}(x)$. In order to compute $\Tilde{g}$, we need an expression for $f_e$, the electron momentum spectrum. Using a thermal distribution,
\begin{equation}
    f_{e,th}(p) = \frac{\beta_{th}}{K_2(\beta_{th})} p^2 e^{-\beta_{th}\sqrt{1+p^2}}
,\end{equation}
yields the correct expression for the SZ effect. Here, $K_x$ is the modified Bessel function and $\beta_{th} = \frac{m_ec^2}{kT_e}$.

Putting this all together, we arrive at a full analytic expression for the suppression factor as a function of temperature in the case of thermal pressure support: this is the curve shown in Fig.~\ref{fig:sup_factor}.

\end{document}